\documentclass[reprint,twocolumn,showpacs]{revtex4}
\pdfoutput=1 
\usepackage{amssymb}
\usepackage{color}

\usepackage{amsthm}
\usepackage{graphicx}
\usepackage{dcolumn}
\usepackage{bm}
\usepackage{subfigure}
\usepackage{amssymb}
\usepackage{verbatim}
\usepackage{amscd}
\usepackage{amssymb}
\usepackage{amsmath, amsfonts}
\usepackage{setspace}
\usepackage[]{graphicx}        
\usepackage{amsthm}
\usepackage{enumerate}
\usepackage{braket}

\theoremstyle{plain}

\theoremstyle{plain}

\theoremstyle{plain}
 
\theoremstyle{plain}

\theoremstyle{remark}

\theoremstyle{conjecture}

\theoremstyle{observation}

\theoremstyle{definition}

\theoremstyle{corollary}

\theoremstyle{definition}

\theoremstyle{definition}

\theoremstyle{assumption}

\theoremstyle{definition}

\theoremstyle{problem}

\theoremstyle{fact}

\begin{document}

\title{Can long-range interactions stabilize quantum memory at nonzero temperature?} 
\author{Olivier Landon-Cardinal}
\affiliation{Institute for Quantum Information and Matter and Walter Burke Institute for Theoretical Physics, California Institute of Technology, Pasadena, California 91125, USA
}
\author{Beni Yoshida}
\affiliation{Institute for Quantum Information and Matter and Walter Burke Institute for Theoretical Physics, California Institute of Technology, Pasadena, California 91125, USA
}
\author{David Poulin}
\affiliation{D\'{e}partement de Physique, Universit\'{e} de Sherbrooke, Qu\'{e}bec, Canada
}
\author{John Preskill}
\affiliation{Institute for Quantum Information and Matter and Walter Burke Institute for Theoretical Physics, California Institute of Technology, Pasadena, California 91125, USA
}
\date{\today}
\begin{abstract}
A two-dimensional topologically ordered quantum memory is well protected against error if the energy gap is large compared to the temperature, but this protection does not improve as the system size increases. We review and critique some recent proposals for improving the memory time by introducing long-range interactions among anyons, noting that instability with respect to small local perturbations of the Hamiltonian is a generic problem for such proposals. We also discuss some broader issues regarding the prospects for scalable quantum memory in two-dimensional systems.
\end{abstract}
\maketitle

\section{Introduction}

Protecting quantum information from damage will be essential for future quantum technologies. Thus, devising a \emph{self-correcting quantum memory}, capable of storing qubits for a very long time without active error correction~\cite{Dennis02, Bacon06, Bravyi09, Beni11}, would be a scientific milestone with far-reaching implications for the scalability of quantum computing and the security of quantum communication protocols. 


We define a self-correcting quantum memory as a quantum many-body system with a local Hamiltonian such that (i) the ground state is degenerate in the limit of a large system, (ii) at a sufficiently small nonzero temperature, quantum information stored in the degenerate ground state is protected from error for a time which grows without bound as the system size increases, (iii) both of these features are stable with respect to generic small local static perturbations of the Hamiltonian.

For a quantum memory to be useful, one will also need a {\em practical} method for storing a quantum state in memory and reading it out, but we do not impose that criterion in our definition. Rather, we regard (ii) as satisfied if it is possible in principle to recover the stored quantum state from a thermally fluctuating state. Topologically ordered systems have a perturbatively stable ground-state degeneracy~\cite{wen1990topological, Bravyi10b}, and would meet all three of our criteria if the topological order persists at nonzero temperature~\cite{Dennis02, Beni11, Hastings11, Castelnovo08, Nussinov09a}.

Known mathematical models of two-dimensional topologically ordered media (such as the toric code~\cite{Kitaev03}) typically do not satisfy (ii)~\cite{Castelnovo07, Nussinov08, Iblisdir09, Alicki09}. The rate per unit area for thermal production of anyons is suppressed by the energy gap, but does not depend on the system size; once produced, thermal anyons can propagate without further energy penalty, quickly destroying the quantum information stored in the ground state~\cite{Bravyi09, Beni10, Haah10, Beni11, Beni10b, Landon-Cardinal13,Pastawski14}. Three-dimensional models have been proposed in which the quantum memory time is enhanced because the propagation of excitations is impeded~\cite{Haah11, Bravyi13, Michnicki12, Michnicki14, Kim12}, but so far the known models of self-correcting quantum memory require at least four spatial dimensions~\cite{Dennis02, Alicki10}. (A recently proposed model based on a low-distortion three-dimensional embedding of a four-dimensional code~\cite{Brell14} has not been shown to be perturbatively stable.) For a recent review of quantum memories at nonzero temperature, see~\cite{BrownReview14}.

To suppress errors arising from the propagation of thermally produced anyons, Hamma \emph{et al.} proposed coupling the toric code to massless bosonic fields, inducing long-range attractive interactions among the anyons~\cite{Hamma09}; however, their scheme requires the coupling strength to diverge with the system size. Pedrocchi {\emph{et al.} proposed an alternative scheme, in which coupling anyons to three-dimensional massless bosons induces an infrared divergent renormalization of the anyon chemical potential, forcing the rate for thermal anyon production to approach zero as the system size grows~\cite{Chesi10b, Pedrocchi12, Pedrocchi13}. Another suggestion is that disorder in a two-dimensional topological medium could localize the anyons and hence extend the memory time~\cite{Stark11,Wootton11}.

In this paper, we emphasize a general issue confronting schemes for stabilizing a two-dimensional quantum memory by coupling to massless bosons --- the criterion (iii) typically fails because generic small local perturbations introduce a mass gap for the bosons and hence compromise the scheme. We consider the model of Pedrocchi \emph{et al.}~\cite{Pedrocchi13} as an instructive example, but our arguments apply more generally, and we also assess some alternative proposals for enhancing memory times. We offer no rigorous arguments excluding self-correcting memory in two or three dimensions, but we hope our discussion helps to clarify the question. Nor do we mean to deny the possibility that clever engineering of interactions or disorder in anyonic systems could substantially enhance the memories times of anyonic systems; our focus is on whether arbitrarily long memory times are achievable as a matter of principle. 

This paper is organized as follows. In Section~\ref{sec:information_loss}, we discuss how Hamiltonian perturbations and thermal fluctuations can corrupt the information stored in a quantum memory, and in Section~\ref{sec:improving} we recall some proposals for enhancing the storage time. In Section~\ref{sec:review}, we discuss the idea of stabilizing a two-dimensional topologically ordered system at nonzero temperature by introducing long-range interactions between anyons, using the ``toric-boson model'' introduced in~\cite{Pedrocchi13} as an explicit example, and in particular we explain how to recover the conclusions of~\cite{Pedrocchi13} using an effective field theory description of the system. In Section~\ref{sec:mass} and \ref{sec:stabilizing}, we argue that the infinite energy barrier of the toric-boson model, and hence its thermal stability, are unstable with respect to generic perturbations of the system's microscopic Hamiltonian.  Section~\ref{sec:discussion} contains some concluding remarks and open questions regarding self-correcting quantum memory.

\section{Logical errors in a quantum memory} \label{sec:information_loss}

To clarify the concept of self-correcting quantum memory, we recall in this Section some ways for a quantum memory to fail, either due to imperfect control of the Hamiltonian or due to thermal fluctuations.

\subsection{Spectral instability}

Desideratum (iii) of our definition of a self-correcting memory requires the degenerate ground space to be stable with respect to small local perturbations of the Hamiltonian. If adding a small local perturbation can lift the degeneracy, then the environment might couple to this local operator, producing fluctuations in the relative phases of energy eigenstates, hence driving decoherence in the energy-eigenstate basis. This failure mechanism relies only on the properties of the nearly-degenerate ground space, not the properties of the excitations supported by the system.

Alternatively, the memory could fail because a weak local perturbation destroys the energy gap separating the ground space from the rest of the energy-eigenstate spectrum. When the energy gap collapses, the structure of the ground space may undergo a qualitative change, allowing the information stored in the ground state to be corrupted. 

We use the term ``spectral instability'' to refer to either of these failure modes. Sufficient conditions for spectral stability, based on topological order, were identified in~\cite{Bravyi10b} and then generalized in~\cite{MZ13}.

\subsection{Thermal relaxation}

Desideratum (ii) of our definition requires the stored quantum information to be well protected from error at a sufficiently small nonzero temperature. Ideally, we would assess whether (ii) is satisfied by modeling the thermal environment, simulating how the stored information is affected by thermal fluctuations, and computing the fidelity of a suitably chosen decoding map. Since this program is quite difficult to carry out, and in any case the results depend on the details of the environment and decoding procedure, we usually assess (ii) using simpler criteria.

\subsubsection{Energy and free energy barriers}\label{sec:energy-barrier}

One way to simplify the story is to imagine that the environment applies a sequence of local quantum operations to the quantum memory. After many such steps, we envision cooling the memory, so that it relaxes back to the ground-state space, and then we compare this final state to the initial state deposited in the memory. We say the error sequence is ``harmful'' if the final state deviates significantly from the initial state.

Each error sequence has an energy height, the maximal value of the expectation value of the Hamiltonian attained during the sequence, and we say that the memory's \emph{energy barrier} is the lowest energy height for any harmful error sequence. Naively, a high energy barrier means a long memory time, because only rare thermal fluctuations will be able to surmount the barrier and damage the stored quantum state.

An Ising ferromagnetic in two or more spatial dimensions is a thermally stable classical memory storing one bit of information in the sign of its magnetization; for the magnetization to flip, thermal fluctuations must produce a large droplet of flipped spins with size comparable to the system size, surmounting an energy barrier scaling like the droplet's surface area. The divergent memory time, though, is not perturbatively stable. A small magnetic field favors one sign of the magnetization; hence bubbles of flipped spins, arising as thermal fluctuations in the disfavored phase at a constant rate per unit time and volume, destroy the stored classical information in a constant time. The three-dimensional toric code, on the other hand, is an example of a self-correcting classical memory, as the one bit detected by its membrane-like logical operator is robust against small local perturbations of the Hamiltonian.

As in the classical case, to ensure a long storage time for a quantum memory we would like the energy barrier to diverge as the system size increases. However, a divergent energy barrier  is not by itself sufficient for thermal stability. Haah's three-dimensional cubic code~\cite{Haah11} has a logarithmically divergent barrier and Michnicki's welded code~\cite{Michnicki12, Michnicki14} has a polynomially divergent barrier, yet in both cases the memory time is bounded above by a constant independent of the system size at any nonzero temperature~\cite{Bravyi13,BrownReview14}.

These examples caution us that even if harmful error sequences are strongly Boltzmann suppressed, the suppression can be overwhelmed if the harmful sequences have high entropy. We should consider not just the energy cost, but rather the free energy cost of encoded errors. In the case of the cubic code, for example, the Boltzmann suppression dominates when the system size is sufficiently small; hence the logarithmically growing energy barrier implies that the memory time $\tau$ increases with system size $L$ as $\tau\sim \exp( c \beta \log L )$, where $c$ is a constant and $\beta$ is the inverse temperature. However once $L$ is larger than a critical system size $L^*\sim e^{c'\beta}$, entropy becomes dominant, and the memory time actually declines with increasing system size. The memory time is optimized for $L\sim L^*$, yielding $\tau < e^{\tilde c\beta^2}$ for some constant $\tilde c$; criterion (ii) for self correction is not satisfied.

Conceivably, entropic effects might enhance the memory time under some circumstances, perhaps allowing the memory to be self correcting even though the energy barrier is finite, as suggested in~\cite{Landon-Cardinal13}. A model in which entropy suppresses harmful error sequences was proposed in~\cite{Brown14} (see also Sec.~\ref{sec:entropically-suppressed}).

\subsubsection{Anyon propagation}\label{sec:anyon-propagation}

Typical topologically ordered systems in two dimensions support anyons --- pointlike quasiparticle excitations obeying exotic statistics. If the system resides on a torus, a logical error occurs when a pair of anyons is created, and then one member of the pair propagates around a homologically nontrivial cycle of the torus before reannihilating with its partner. Thus the energy barrier is the (constant) energy cost of creating and separating a pair of anyons. 

At nonzero temperature, thermal fluctuations produce a nonvanishing density of anyons per unit area. We can imagine decoding the state by bringing pairs of anyons together to annihilate, returning the system to a ground state. Starting with an initial noisy encoded state, we may allow the system to evolve for a while in contact with a heat bath, before applying the decoding map to the final configuration. By combining the anyon world lines arising from the decoding of the initial state, the decoding of the final state, and the thermally induced anyon propagation, creation, and annihilation between the initial and final time, we obtain a cycle, a set of closed loops. If this cycle is homologically nontrivial, a logical (encoded) error occurs. 

When the anyons propagate a distance comparable to the typical (constant) distance between particles in the thermal anyon gas, a logical error will occur with nonnegligible probability. Therefore, we do not normally expect a two-dimensional quantum memory to be self correcting, unless supplemented by an additional mechanism to impede the propagation or creation of anyons. 

The three-dimensional version of the toric code also has pointlike ``anyon'' excitations, and a logical error occurs when an anyon propagates across the system. Therefore this code, too, has a constant energy barrier and is not self correcting. The cubic code~\cite{Haah11} has pointlike excitations, but reaching a configuration in which one particle is separated from all others by a distance at least $R$ requires energy scaling like $\log R$; thus thermal particle propagation is suppressed and the memory has a logarithmically growing energy barrier. Nevertheless, as noted in Sec.~\ref{sec:energy-barrier}, the free energy cost of a logical error becomes independent of system size for a sufficiently large system at a fixed temperature.

In the three-dimensional welded code~\cite{Michnicki12, Michnicki14}, $v > 2$ three-dimensional toric code subblocks are sewn together at each weld. Anyons propagate freely within each subblock, but for an anyon to pass through a weld from one subblock to another, at least $v-2$ additional anyons must be created. Therefore the energy barrier is high if the code has many such subblocks. However, logical errors can arise at nonzero temperature from a process in which $v$ thermal anyons propagating in distinct subblocks meet at a weld and fuse to the vacuum~\cite{BrownReview14}. This process is entropically enhanced because the fusion can occur at any point along the weld; the resulting logical error rate is independent of system size and therefore the code is not self correcting.

\section{Improving the storage time}\label{sec:improving}

Having discussed how a quantum memory can fail, we now briefly consider some possible ways to improve the stability of a quantum memory and hence enhance its storage time.

\subsection{Anyon localization}

Even for a topologically ordered system with a spectral gap, in which small local perturbations split the ground state degeneracy by an amount which is exponentially suppressed for a large system size, a quantum memory may fail due to small local perturbations of the Hamiltonian. The problem is that, although information encoded in the ground space may be well protected in principle, if the perturbation is unknown it may be difficult to prepare precisely the ground state of the perturbed system. 

For example, if the ideal system has the toric code Hamiltonian, we might write to the memory by preparing one of the toric code ground states. But if the Hamiltonian is slightly deformed by a weak perturbation, the ground state of the ideal Hamiltonian is a coherent superposition of excited states of the perturbed Hamiltonian; this superposition dephases quickly, destroying the encoded information in a constant time independent of system size~\cite{Pastawski09}. If the perturbation were exactly known, then in principle we could decode the state by running the Hamiltonian evolution backward in time. But if (more realistically) the perturbation is unknown, then an uncorrectable logical error is likely to occur. This error mechanism operates even at zero temperature; the error arises not from thermal fluctuations but rather from imperfect control of the system's Hamiltonian.

One way to evade this problem is to respond to the deformation of the Hamiltonian by adjusting the procedure for encoding and decoding the state. Under a generic perturbation, the static pointlike anyons of the exact toric code  become propagating dressed anyons with nonzero size. Using a physical device that detects these dressed anyons, we can prepare a ground state of the perturbed system, which will store quantum information reliably at zero temperature. In principle we could read out the encoded state at a later time using dressed string operators.

Alternatively, these zero-temperature errors can be suppressed by introducing disorder in the system Hamiltonian, ensuring that in low-lying excited states the anyons are spatially localized rather than freely propagating~\cite{Wootton11}. In fact, localization may suffice to make the memory time exponentially large in the system size, when the errors are due to unknown Hamiltonian perturbations~\cite{Bravyi12b,Stark11}. However, at any nonzero temperature, {\em single-particle} localization does not prevent anyons from diffusing at a nonzero rate due to thermally activated hopping through the disordered system. Thus at nonzero temperature, though localization may enhance the memory time by a constant factor, we still expect the logical errors to occur after a constant time independent of system size --- the memory is not self correcting.

We note that a disordered topologically ordered system may exhibit {\em many-body localization}, in which anyons are spatially localized even in highly excited states of the disordered Hamiltonian~\cite{Huse1304.1158,Bauer1306.5753}. However, many-body localization is a property of a perfectly isolated Hamiltonian system. Even a many-body-localized two-dimensional topological phase would fail to be self correcting, because anyons would still be able to diffuse when the system is in contact with a thermal bath. 

For the case of a two-dimensional commuting Hamiltonian like the disordered toric code model, the finding that topological order is destroyed at nonzero temperature can be formulated rigorously. Hastings~\cite{Hastings11} showed that the Gibbs state of a (not necessarily translation-invariant) 2D commuting local Hamiltonian can be (approximately) transformed to a product state by a constant-depth local quantum circuit. To analyze the memory time, we can draw guidance from studies of the disordered 1D Ising model, whose stability properties are similar to those of the 2D toric code. For a one-dimensional classical spin system with Hamiltonian 
\begin{align}
H_{Ising}= -\sum_{i}J_{i}Z_{i}Z_{i+1} \label{eq:disorder}
\end{align}
where the nearest-neighbor couplings $\{J_{i}\}$ are positive but otherwise arbitrary, the memory time $\tau$ is independent of system size, scaling like $\tau \sim \exp(2\beta \bar J)$ where $\bar J\equiv(J_{min}+J_{max})/2$ is the average of the minimal and maximal values of the coupling~\cite{Droz86}. Thus the memory time is not significantly enhanced in the regime $J_{max}\gg J_{min}$ where defects are strongly localized. The conclusion $\tau = O(1)$ also applies to the (disordered) toric code, subject to a master equation whose jump operators are Pauli operators; in that case the jumps transform energy eigenstates to energy eigenstates, so that the convergence to a fixed point can be analyzed using classical methods~\cite{Alicki09}.

\subsection{Entropically suppressed propagation}\label{sec:entropically-suppressed}
Brown \emph{et al.} described an anyon model in which anyon propagation is impeded by entropic considerations~\cite{Brown14}. In their model, a two-dimensional topologically ordered medium is divided into domains separated by defect lines. Each domain supports both heavy and light anyons, where the decay of a heavy anyon into two light anyons is allowed by both kinematics and the fusion rules. Furthermore, upon crossing a defect line, a heavy anyons is transformed into a light one, and a light one is transformed into a heavy one. 

Propagation of light anyons across defect lines is suppressed at low temperature, because an energy boost is needed to cross to the other side. Furthermore, it is entropically favored for heavy anyons to decay to pairs of light ones. Therefore, anyons cannot easily travel across the system and the memory time is correspondingly enhanced. Numerical evidence suggests that if the system size is properly optimized, the memory time $\tau$ scales with the inverse temperature $\beta$ as $\tau \sim e^{c \beta^2}$, just as in the three-dimensional cubic code, even though the energy barrier is constant rather than logarithmically growing. This is a notable improvement compared to the scaling $\tau\sim e^{c \beta}$ in the toric code. Though not self correcting, this system illustrates how in some cases favorable entropic effects can significantly enhance the memory time.

\subsection{Long-range anyon interactions}

Since logical errors in a two-dimensional topologically ordered quantum memory arise from anyon propagation, any mechanism that impedes anyon propagation can in principle enhance the memory time. Aside from localization due to disorder as discussed above, there is another mechanism that could limit how far anyons can travel --- attractive interactions between anyons. The interactions may impose a hefty energy penalty on harmful error sequences, encouraging confinement of the anyons to small dilute clusters which do not produce logical errors. For the system to be truly self correcting, though, these interactions must have infinite range. Otherwise only a constant amount of energy would be needed for an anyon to break away from a cluster, leaving it free to propagate unmolested by attractive forces applied by other anyons, thus allowing logical errors to occur at a constant rate at any nonzero temperature. 

For a system with a local Hamiltonian, long-range forces arise from the exchange of gapless excitations. Hamma \emph{et al.}~\cite{Hamma09} proposed a model in which the anyon density provides a source for a massless scalar field. Scalar exchange produces long-range attractive interactions between anyons, but each anyon also has a logarithmically infrared divergent negative self energy, rendering the system unstable. The divergent self-energy can be cancelled by introducing other fields with carefully tuned couplings, but then the Hamiltonian contains terms with infrared divergent coefficients. 


Improving on \cite{Hamma09}, Pedrocchi \emph{et al.} proposed an alternative model with a local Hamiltonian and finite couplings, in which the anyonic chemical potential is infrared divergent ~\cite{Pedrocchi13}. Thus the energy barrier, the cost of creating a distantly separated anyon pair, diverges with system size, as does the memory time. We review this scheme in Sec.~\ref{sec:review}. It works, if the Hamiltonian is carefully tuned. But as we explain in Sec.~\ref{sec:mass} it does not meet our criteria for self correction, because the divergent energy barrier is unstable with respect to generic small perturbations of the local Hamiltonian. 

We also note that schemes for ``active'' rather than ``passive'' error correction have been proposed, inspired by the idea that long-range attractive interactions might impede the propagation of diffusing anyons \cite{harrington2004analysis}. In a scheme recently discussed in \cite{Herold1406.2338}, the anyon motion is guided by a classical field, which is regularly updated using a local evolution rule as the anyons propagate. Such active error correction schemes, though effective and interesting, do not meet our criteria for self correction for several reasons. For one thing, to achieve an arbitrarily low logical error rate, the scheme requires a ``speed of light'' (rate of information propagation in the classical field) which diverges relative to the ``speed of sound'' (rate of anyon motion), a requirement not satisfied by a system with a local Hamiltonian and bounded couplings. Furthermore, the dissipative dynamics of the field which controls the anyons may not be realizable by any Hamiltonian system in contact with a thermal bath. 

\section{Toric code coupled to a scalar field}\label{sec:review}

\subsection{Toric code}
To prepare for our discussion of models with long-range anyon interactions, let's first recall Wen's formulation~\cite{Wen03} of Kitaev's toric code model~\cite{Kitaev03}. Consider an $L\times L$ square lattice with one qubit per site, where $L$ is an even integer. The Hamiltonian is
\begin{align}
H_{Toric} = -J \sum_{r} W_{r}, \quad W_{r}=X_{r,1}Z_{r,2}X_{r,3}Z_{r,4}, \label{eq:Toric}
\end{align}
where $r$ labels lattice plaquettes and $W_{r}$ acts on the four qubits contained in plaquette $r$  ($X$ and $Z$ denote Pauli matrices). On the torus ({\em i.e.} a square with periodic boundary conditions), the model has four degenerate ground states which can encode two qubits. 

The energy cost of creating an anyon pair is $4J$, and the anyons are noninteracting. A logical error occurs if a pair is created and one anyon propagates around a nontrivial cycle of the torus before the anyons reannhihilate. Therefore the (constant) energy barrier is $4J$. The typical separation between particles in the thermal anyon gas, and hence also the memory time,  scale like the inverse Boltzmann factor $e^{2\beta J}$.


\subsection{Coupling the toric code to massless bosons}\label{subsec:coupling}

Now, following Pedrocchi \emph{et al.}~\cite{Pedrocchi13}, we consider coupling the two-dimensional toric code to an auxiliary three-dimensional bosonic system which induces long-range attractive forces among the plaquette variables. The Hamiltonian proposed by \cite{Pedrocchi13} is
\begin{equation}\label{eq:Pedrocchi-Hamiltonian}
\begin{split}
&H = -A \sum_{r}W_{r}(a_{r}+a^{\dagger}_{r}) + H_{boson},\\ 
&H_{boson} = \sum_{q}\epsilon_{q}a^{\dagger}_{q}a_{q},\\ 
&a_q = \frac{1}{\sqrt{N}}\sum_s a_s e^{iqs}.\\ 
\end{split}
\end{equation}
Here the index $r$ labels the sites of the two-dimensional topologically ordered medium and $s$ labels the $N$ sites of a three-dimensional lattice which contains the two-dimensional lattice; $a_r$ and $a_r^\dagger$ are the annihilation and creation operators for bosonic modes residing at the location of plaquette operator $W_r$, and  $q$ labels the discrete Fourier modes for a bosonic field defined on the three-dimensional lattice. We assume $\epsilon_q > 0$, because a bosonic mode with negative $\epsilon_q$ would be unstable. The precise behavior of $\epsilon_q$ for large $q$ does not matter for analyzing the long-distance physics of the model, but the behavior of $\epsilon_q$ for small $q$ does matter. We assume a gapless spectrum for the Fourier modes with $\epsilon_{q}\sim q^2$ for small $q$. We will call the Hamiltonian in Eq.~(\ref{eq:Pedrocchi-Hamiltonian}) the ``toric-boson model.''

Coupling anyons to the three-dimensional massless boson field produces an attractive force between anyons, with potential energy inversely proportional to separation. After a polaron transformation (\emph{i.e.}, completing the square) the Hamiltonian may be rewritten as~\cite{Pedrocchi13}
\begin{align}
H = - \sum_{r,r'}J_{r,r'}W_{r}W_{r'} + \sum_{q}\epsilon_{q}\tilde{a}^{\dagger}_{q}\tilde{a}_{q},  \label{eq:Toric_Boson}
\end{align}
where
\begin{align}\label{eq:Jrrprime}
J_{r,r'} = \frac{A^2}{N}\sum_{q} \frac{1}{\epsilon_{q}} e^{iq(r-r')} \sim  \frac{A^2}{|r-r'|};
\end{align}
here $N$ is the number of lattice sites in the three-dimensional bosonic system, 
\begin{equation}
\tilde{a}_{q} = a_{q}-\frac{A}{\epsilon_{q}\sqrt{N}} \sum_{r} W_{r}e^{iqr},\end{equation}
and the conclusion $J_{r,r,'} \sim {A^2}/{|r-r'|}$ applies for asymptotically large separation. Note that $\tilde{a}_{q}$ and $\tilde{a}_{q}^{\dagger}$ both obey ordinary bosonic commutation relations because the $W_{r}$'s residing at distinct plaquettes are commuting operators. The long-range behavior of the potential energy, scaling like $\sim 1/R$ where $R$ is the separation, arises from the pole in $1/\epsilon_{q}$, which dominates the Fourier integral for large $R$.

Because the Hamiltonian in Eq.~(\ref{eq:Toric_Boson}) is the sum of an anyonic part $H_{A}$ and bosonic part $H_{B}$, which are completely decoupled, the energy eigenstates can be written as $|\psi\rangle = |\psi_{qubit}\rangle \otimes |\phi_{boson}\rangle$, where  $|\psi_{qubit}\rangle$ is an eigenstate of $H_A$ and $|\phi_{boson}\rangle$ is an eigenstate of $H_B$. Since $J_{r,r'}>0$, the interactions between plaquette variables in Eq.~(\ref{eq:Toric_Boson}) are ``ferromagnetic,'' favoring alignment. $H_A$ has a $\mathbb{Z}_{2}$ global symmetry, under which $W_r$ changes sign, which is spontaneously broken at zero temperature --- there is one ground state sector with $W_{r}=1$ for all $r$ and another sector with $W_r = -1$ for all $r$, where both sectors have a topological degeneracy as well. The spontaneous breakdown of the global $\mathbb{Z}_{2}$ symmetry persists at sufficiently small nonzero temperature, and we may assume without loss of generality that the thermal expectation value $\langle W_r\rangle$ is positive .



As emphasized in~\cite{Pedrocchi13}, the crucial property of Eq.~(\ref{eq:Toric_Boson}) is that the energy cost of creating a single isolated anyon is infrared divergent, \emph{i.e.}, increases with the system size. 
To compute this energy cost, we follow~\cite{Pedrocchi13} and suppose that the bosonic bath is at any moment in a Gibbs state with respect to the effective Hamiltonian $\langle \alpha | H | \alpha \rangle $, where $|\alpha\rangle$ is the instantaneous anyonic state, \emph{i.e.}, an eigenstate of all operators $W_r$. In that case, this energy cost of creating a single isolated anyon only depends on the anyonic part $H_A$ of the Hamiltonian in Eq.~\eqref{eq:Toric_Boson} (see~\cite{Pedrocchi13} for the detailed calculation). The divergence arises because each plaquette interacts with many other distantly separated plaquettes on the two-dimensional lattice; flipping the sign of one $W_r$ relative to all the others increases the potential energy by an amount
\begin{align} \label{eq:chem-pot-linear}
\mu \sim \sum_{r'}J_{r,r'} \sim O(L).
\end{align}
This effective chemical potential $\mu$ is $O(L)$ because we sum the $1/R$ interaction energy over the $L\times L$ lattice. In contrast to the standard toric code, where the energy of an isolated anyon is $2J$ and the thermal density of anyons at inverse temperature $\beta$ scales like $e^{2\beta J}$, in the model Eq.~(\ref{eq:Toric_Boson}) the thermal density scales like $e^{c \beta L}$ (where $c$ is a constant), and correspondingly the memory time grows exponentially with $L$ at any nonzero temperature. The infrared divergent chemical potential and the resulting self-correcting behavior of Eq.~(\ref{eq:Toric_Boson}) were the main conclusions of \cite{Pedrocchi13}. In fact, the infinite chemical potential means that the spontaneous breakdown of the $\mathbb{Z}_{2}$ global symmetry, and hence the self-correcting behavior, hold at arbitrarily high temperature.


\subsection{Derivation using effective field theory}\label{subsec:EFT}

Because the conclusion $\mu = O(L)$ follows from only the long-distance properties of the Hamiltonian Eq.~(\ref{eq:Toric_Boson}), it can be recovered using an effective field theory description which captures the infrared behavior of the model. The effective Hamiltonian becomes
\begin{equation}
H_{EFT}=\tfrac{1}{2}\int d^{3}x\:\left(\nabla\phi\right)^{2}-A\int d^{2}y\: w(y)\phi(y)+ \cdots, \label{eq:QFT-Hamiltonian}
\end{equation}
where $\phi(x)$ is a suitably normalized scalar field. The $\tfrac{1}{2}\left(\nabla\phi\right)^{2}$ term in the Hamiltonian density arises from the dispersion relation $\epsilon_{q}\sim q^2$ for small $q$; higher-order terms in $q$ correspond to terms with more derivatives in $H_{EFT}$, which are not shown because they produce only small corrections for long-wavelength modes of the scalar field. The source $w(y)$ describes the anyon configuration. To be compatible with the conventions of Sec.~\ref{subsec:coupling}, we assume $w(y)$ takes the positive value $w_0>0$ in the ground state of the two-dimensional topological medium, and has the value $w(y) = w_0 - 2 n(y)$ when anyons are present, where $n(y)$ is a suitably smoothed and normalized anyon density. We assume $w(y) = 0$ outside the two-dimensional medium.


  
The energy is minimized by a solution to the static classical field equation, which is Poisson's equation 
\begin{equation}
\nabla^{2}\phi(x) =-Aw(x).\label{eq:field_equation}
\end{equation}
The solution is 
\begin{equation}
\phi(x)=A\int d^{2}y\left(\tfrac{1}{-\nabla^{2}}\right)_{x,y}w(y)+\phi_{0}(x)\label{eq:field_expression}
\end{equation}
where $\phi_0(x)$ is a solution to the homogeneous field equation, and the Green function $\tfrac{1}{-\nabla^{2}}$ is determined by imposing appropriate boundary conditions. We assume that the two-dimensional anyonic source for the three-dimensional scalar field $\phi$ has bounded support, and that $\phi$ vanishes at spatial infinity.

The boundary condition justifies an integration by parts with no surface term, and using the field equation we may obtain an alternative expression for the minimal energy of the field configuration in the presence of an anyon source:
\begin{eqnarray}
E[w] & = & -\int d^{3}x\,\tfrac{1}{2}\phi\nabla^{2}\phi-A\int d^{2}y\: w(y)\phi(y)\nonumber\\
 & = & \tfrac{A}{2}\int d^{3}x\, w(x)\phi(x)-A\int d^{2}y\: w(y)\phi(y)\nonumber\\
 & = & -\tfrac{A}{2}\int d^{2}x\: w(x)\phi(x) 
\end{eqnarray}
with the integral now supported only on the two-dimensional surface where the source field $w$ is nonzero. Inserting Eq.~(\ref{eq:field_expression}) into this expression, one finds
\begin{equation}\label{eq:E(w)-EFT}
E[w]=-\tfrac{A^{2}}{2}\int d^{2}x\, d^{2}y\: w(x)\left(\tfrac{1}{-\nabla^{2}}\right)_{x,y}w(y).
\end{equation}
The boundary condition determines the Green function to be 
\begin{equation}
\left(\tfrac{1}{-\nabla^{2}}\right)_{x,y}=\frac{1}{4\pi \left|x-y\right|},
\end{equation}
and so we obtain
\begin{equation}
E[w]=-\tfrac{1}{2}\int d^{2}x\, d^{2}y\: w(x)J(x,y)w(y),
\end{equation}
where
\begin{equation}
J(x,y)= \frac{A^2}{4 \pi|x-y|},
\end{equation}
as in Eq.~(\ref{eq:Jrrprime}).

To compute the effective anyon chemical potential, we write $w(y)=w_0-2n(y)$, finding that $E[w]$ contains a term linear in the anyon density 
\begin{equation}
\int d^{2}x\: \mu(x) n(x)
\end{equation}
where
\begin{equation}
\mu(x) = 2w_0\int d^{2}y\, J(x,y).
\end{equation}
Assuming for simplicity that $x$ is the center of a disk with radius $L/2$, we obtain
\begin{eqnarray}\label{eq:chemical_potential}
\mu & = & 2w_0 A^{2}\int_{0}^{2\pi}d\theta\int_{0}^{L/2}rdr\tfrac{1}{4\pi r}=\tfrac{w_0A^{2}}{2}L,
\end{eqnarray}
recovering the linear divergence found in Eq.~(\ref{eq:chem-pot-linear}).

\section{Perturbative instability of massless bosons}\label{sec:mass}

Having seen that the model in Eq.~(\ref{eq:Pedrocchi-Hamiltonian}) has an infrared divergent energy barrier, and hence an exponentially increasing quantum storage time at sufficiently small nonzero temperature, we next consider the stability of the energy barrier with respect to small perturbations of the system's local Hamiltonian. For assessing perturbative stability, it is important to note that the divergent anyon chemical potential arises because the response of the scalar field to the anyon vacuum is infrared divergent. A small perturbation of the Hamiltonian can tame this divergence, rendering the energy barrier finite.

The bosonic ground state $|\phi_{boson}\rangle$ of Eq.~(\ref{eq:Toric_Boson}) satisfies 
\begin{equation}
\tilde{a}_{q}|\phi_{boson}\rangle = 0
\end{equation}
for all $q$, which implies
\begin{align}
\langle a_{r}\rangle_{0} \sim \frac{A}{N}\sum_{q}\frac{1}{\epsilon_{q}} \sum_{r'} e^{iq(r'-r)}   =O(L)
\end{align}
where $\langle a_{r}\rangle_{0}$ denotes the ground-state expectation value, and we have substituted $W_{r'} = 1$. The bosonic mode $a_r$ has an unbounded occupation number in the anyon ground state, which is why the cost of creating an isolated anyon at site $r$ is divergent.

Additional terms in the bosonic Hamiltonian, which are generically present, will prevent the occupation number from diverging. Suppose, for example, the Hamiltonian contains a term
\begin{align}
V = \epsilon \sum_{r} n_{r} = \epsilon \sum_q a_q^\dagger a_q, \label{eq:perturbation}
\end{align}
where $n_r= a_r^\dagger a_r$ is the number operator for the mode localized at site $r$. We may assume $\epsilon > 0$ because otherwise the bosonic field would be unstable, an indication that we are expanding around the wrong bosonic vacuum. Now the perturbation $V$, which disfavors a large occupation number, competes with the Hamiltonian Eq.~(\ref{eq:Toric_Boson}), and the ground-state expectation value of $a_r$ takes the finite value
\begin{align}\label{eq:finite-mass-a}
\langle a_{r}\rangle_{0} \sim \frac{A}{N}\sum_{q}\frac{1}{\epsilon+\epsilon_{q}} \sum_{r'} e^{iq(r'-r)}   = O(\epsilon^{-1/2}),
\end{align}
because the perturbation removes the pole at $q=0$. 

For this particular perturbation the bosonic Hamiltonian is quadratic and therefore exactly solvable, but the conclusion that a generic perturbation removes the infrared divergence holds more generally. In the effective field theory language, the solution to the static field equation Eq.(\ref{eq:field_expression}) becomes
\begin{eqnarray}
&\phi(x)=A\int d^{2}y\left(\tfrac{1}{-\nabla^{2}}\right)_{x,y}w(y)\nonumber\\
&= Aw_0\int d^{2}y \frac{1}{4\pi |x-y|} = O(L)
\end{eqnarray}
in the anyonic vacuum such that $w(y) = w_0$ in the two-dimensional topological medium and $w(y) = 0$ elsewhere. The infrared divergent behavior of  $\phi(x)$ arises because the scalar field is exactly massless in Eq.~(\ref{eq:QFT-Hamiltonian}). Generically a nonzero mass term will be present, perturbing the Hamiltonian according to
\begin{align}
H_{EFT}\to H_{EFT} + \int d^{3}x\:  \tfrac{1}{2} m^2\phi^2.
\end{align}
We may assume $m^2 > 0$, because otherwise the scalar field would be unstable, indicating that we are expanding around the wrong bosonic vacuum. The static vacuum solution becomes
\begin{eqnarray}\label{eq:finite-mass-phi}
\phi(x)=Aw_0\int d^{2}y\left(\tfrac{1}{-\nabla^{2}+m^2}\right)_{x,y}=O(m^{-1});
\end{eqnarray}
now the $y$ integral converges because the Green function decays like $e^{-m|x-y|}$ for large separation.

The effective field theory approach is useful for analyzing the energy barrier of a quantum memory because whether the energy barrier diverges or not hinges on the behavior of the model in the far infrared, where the field theory approximation should be reliable. The degrees of freedom described by the field theory may be \emph{emergent}, not corresponding in any very direct with the microscopic degrees of freedom of the underlying Hamiltonian model. But just as the divergent energy barrier in the model Eq.~(\ref{eq:QFT-Hamiltonian}) is realized only when the free parameter $m^2$ of the effective field theory is tuned to zero, we may anticipate that a divergent energy barrier arises from the exchange of massless bosons only on (at best) a codimension-one surface in the space of all Hamiltonian models. A generic perturbation moves the Hamiltonian off this surface, destroying the infinite energy barrier. To evade this conclusion, we need models in which bosons with the appropriate couplings remain gapless when the microscopic Hamiltonian is deformed.

Not just the infinite energy barrier but also the topological order of the two-dimensional medium might be threatened by small perturbations of the Hamiltonian. In addition to coupling to plaquette variables, the scalar field could couple to individual qubits; for example the perturbed Hamiltonian might contain terms such as
\begin{equation}\label{eq:B-field-pert}
V = \epsilon' \sum_r (a_r + a_r^\dagger)(X_{r,1}+X_{r,2}+X_{r,3}+X_{r,4}),
\end{equation}
If in the ground state the expectation value of the field scales like $\langle a_r \rangle_0\sim m^{-1} \sim \epsilon^{-1/2}$ as in Eq.~(\ref{eq:finite-mass-a}) and (\ref{eq:finite-mass-phi}), then an effective magnetic field 
\begin{equation}\label{eq:B-field-scaling}
B\sim \frac{\epsilon'}{\sqrt{\epsilon}}
\end{equation} 
is applied to the qubits. Although enhanced by the large strength of the field, this magnetic field is still perturbatively small compared to the energy cost $\sim 1/\sqrt{\epsilon}$ of creating an isolated anyon, so it is plausible that the topological order would survive. In any case, we expect the topological order to persist within a region of finite volume in the space of Hamiltonians.

\section{Stabilizing massless bosons}\label{sec:stabilizing}

The scheme of Pedrocchi \emph{et al.}~\cite{Pedrocchi13} for stabilizing a quantum memory requires massless bosonic fields, and we have argued that under generic conditions bosonic fields acquire mass unless the Hamiltonian is tuned to a codimension-one surface. But there are important exceptions to this rule, and we should consider whether these exceptions can be exploited by quantum engineers. In particular, the masslessness of Goldstone bosons and gauge bosons follows from general symmetry principles.

\subsection{Goldstone bosons}

A spontaneously broken exact continuous symmetry is always associated with an exactly massless Goldstone boson, whose masslessness is preserved by any perturbation that respects the symmetry (and also maintains its spontaneous breakdown). If $\phi(x)$ is a Goldstone field, then the symmetry acts according to
\begin{equation}
\phi(x) \mapsto \phi(x) + \mbox{constant},
\end{equation}
disallowing a mass term $\tfrac{1}{2}m^2\phi^2$ in the effective Hamiltonian. 

The symmetry not only forbids a mass term for the Goldstone boson, but also constrains its coupling to other fields. For example, a term $-A\phi(x)w(x)$ describing the coupling of the Goldstone field to the anyonic density $w(x)$ is not allowed, but a derivative coupling $-A\nabla \phi(x) w(x)$ is compatible with the symmetry, and can be reexpressed as $A\phi(x)\nabla w(x)$ after integrating by parts. (This term violates rotational invariance, but could be allowed in an anisotropic system.) Then, arguing as in Sec.~\ref{subsec:EFT}, we find in place of Eq.~(\ref{eq:E(w)-EFT}) that the energy of the bosonic ground state in the presence of a source becomes (in $D$ spatial dimensions)
\begin{eqnarray}
&E[w]=-\tfrac{A^{2}}{2}\int d^{D}x\, d^{D}y\: \nabla w(x)\left(\tfrac{1}{-\nabla^{2}}\right)_{x,y}\nabla w(y)\nonumber\\
&= -\tfrac{A^{2}}{2}\int d^{D}x\, w(x)^2,
\end{eqnarray}
the infrared-finite integral of a local source-dependent energy density. Because they are derivatively coupled, the exchange of Goldstone bosons generates short-range contact interactions between anyons rather than a long-range forces. 

Perturbations that explicitly break the continuous symmetry would gap out the Goldstone bosons in any case. But even if we are willing to impose a symmetry that protects the massless Goldstone bosons, there would be no resulting divergent energy barrier to suppress logical errors.

\subsection{Gauge bosons}

In some cases, exactly massless deconfined $U(1)$ gauge bosons can emerge from an underlying spin model with no fundamental gauge symmetry. From an effective field theory viewpoint, since the gauge symmetry is not an exact feature of the underlying microscopic Hamiltonian, nothing seems to forbid adding a gauge non-invariant term to the Hamiltonian of the long-distance effective field theory, which would gap out the gauge bosons. But if the microscopic local Hamiltonian energetically favors a sector of the theory which satisfies gauge constraints, these constraints could be robust against generic small perturbations of the microscopic Hamiltonian~\cite{hastings-cond-mat-0503554}. A model with such features can support a stable ``Coulomb phase,'' where a ``photon'' remains exactly massless without any need to impose symmetries or carefully tune the Hamiltonian. 

Emergent photons can mediate long-range forces between charged particles, which are attractive for particles of opposite charge, and repulsive for particles of like charge. Chesi \emph{et al.} suggested that long-range repulsive interactions among anyons in a two-dimensional medium could produce a divergent free energy barrier, because for a nonzero anyon density the cost of introducing an additional isolated anyon would diverge with system size~\cite{Chesi10b}; thus the anyon density should approach zero in the thermodynamic limit.  

However, anyon models in which gauge-mediated interactions are repulsive for each anyon pair do not seem to be self consistent. Because the photon couples to a locally conserved charge, anyon pairs which can be locally created from the vacuum must have opposite ``electric charges.'' At nonzero temperature, anyons coupled to emergent photons would form a plasma with charge screening and no divergent energy barrier 

It may be possible in principle for anyons to carry dipole moments which couple to emergent photons. In that case, if the gauge field is three-dimensional there would be a long-range interaction with potential energy falling of with distance $R$ like $1/R^3$. Even if in a suitably engineered Hamiltonian the 3D gauge field couples to the plaquette variables of the 2D toric code, these interactions decay too rapidly to produce an infrared divergent energy barrier. 

For the 3D toric code, a logarithmically divergent chemical potential $\mu \sim \log L$ due to dipolar interactions is conceivable. It's not clear, however, how to arrange the needed coupling between the topological medium and the gauge theory; see Sec.~\ref{sec:perturbation} for further discussion.

\subsection{Other long-range forces}

Massless scalars can also be stabilized by imposing both supersymmetry and chiral symmetry -- the chiral symmetry enforces the masslessness of a fermion species, and the supersymmetry requires the scalar and fermion to be degenerate. Massless spin-two bosons (gravitons) can be stabilized by general covariance. If we could find spin models in which supersymmetry and/or general covariance are emergent and robust with respect to generic local perturbations of the microscopic Hamiltonian, that would be very interesting from the perspective of fundamental physics, apart from the potential applications to self-correcting quantum memory!

Putting aside the microscopic origin and perturbative stability of the long-range interactions, we may consider the consequences of long-range forces between ``anyons'' in $D$ dimensions described by Eq.~(\ref{eq:Toric_Boson}), where 
\begin{align}
J_{r,r'} \sim \frac{1}{|r-r'|^{\alpha}}.
\end{align}  
For $0<\alpha<D$, the cost of creating an isolated anyonic excitation diverges with the system size, so the model has a quantum memory time which increases rapidly with system size, even at arbitrarily high temperature. The precise scaling of the lifetime of the memory with system size is discussed in~\cite{Wootton13}.
For $\alpha >D$, energy cost remains finite, and the memory time is bounded above by a constant at any nonzero temperature. The case $\alpha=D$ is marginal, with a logarithmically divergent energy barrier. 

By similar reasoning, the one-dimensional Ising model with $0<\alpha<1$ is a good classical memory at arbitrarily high temperature. It is also known that this system has a finite-temperature phase transition for $1<\alpha < 2$, and is magnetically disordered at any nonzero temperature for $\alpha > 2$~\cite{Dyson69,Mukamel09}.


\section{Discussion}\label{sec:discussion}

The main thrust of this paper is contained in Sec.~\ref{sec:review}-\ref{sec:stabilizing}, where we have described schemes for stabilizing a two-dimensional topologically ordered quantum memory at nonzero temperature using long-range interactions, and have argued that such schemes have a significant and generic drawback --- the Hamiltonian needs to be precisely tuned for the memory time to scale favorably with the system size. Our arguments are incomplete, and leave many open questions, including the important and enticing question whether self-correcting quantum memory is achievable in a suitably engineered three-dimensional system. This concluding section contains a few questions and remarks which may help to guide future research on self-correcting quantum memory.



\subsection{Quasi-topological phases}

In mathematical models of topologically ordered systems, we usually assume the system has an energy gap separating the topologically degenerate ground state from the rest of the energy spectrum. Then we can imagine integrating out all quasiparticle excitations to obtain an effective theory with no propagating excitations, a topological quantum field theory (TQFT) which accurately describes the physics of the system up to corrections which are exponentially small in the system size. But strictly speaking this picture does not necessarily apply to the topologically ordered systems observed in the laboratory, which may support gapless photon or phonon excitations. Furthermore, disordered systems may have many low-lying localized bulk quasiparticle excitations with energy well below the topological gap. 

Bonderson and Nayak proposed the term \emph{quasi-topological phase} for a system that has some of the features of a topological phase, but also has gapless excitations and therefore cannot be precisely described at low energy by a TQFT ~\cite{Bonderson13}. For example, Laughlin states are quasi-topological; their fractionally charged quasiparticles couple to the massless photon, and correspondingly the topological degeneracy on the torus is split by an amount that scales polynomially rather than exponentially with the system size. Furthermore, the anyon braiding statistics is modified by nonuniversal dynamical phases which are not described by the TQFT. Bonderson and Nayak also proposed the term \emph{strong quasi-topological phase} for a system in which some sector of the theory decouples from the gapless excitations and can be accurately described by a TQFT. For example, the Moore-Read state has a strongly quasi-topological sector. Although the Ising anyons carry electric charges which produce non-topological exchange phases, the distinguishable fusion channels for a system of many Ising anyons span a topologically protected Hilbert space well described by the Ising TQFT.

The toric-boson model, the toric code coupled to an exactly massless scalar field as in Eq.~(\ref{eq:Toric_Boson}), is a strong quasi-topological phase. One might have thought that, just as distantly separated anyons have a potential energy scaling algebraically with their separation, so large punctures on the lattice, which we may think of as fattened anyons, should also have an algebraically decaying interaction, and corresponding the splitting of the degeneracy on the torus should be algebraically decaying as for Laughlin states. But in the exactly solvable toric-boson model, the topological degeneracy on the torus is exact. Deforming the model with a small local perturbation might conceivably destroy the topological order as discussed in Sec.~\ref{sec:mass}, but in the ``weak magnetic field'' regime of Eq.~(\ref{eq:B-field-pert}) and (\ref{eq:B-field-scaling}) the perturbation may produce only an exponentially small splitting of the ground state degeneracy, just as for the toric code without long-range interactions. 







\subsection{Topological order at low temperature in a finite system}

Our focus throughout this paper has been on the scaling of the memory time with system size, and in particular on whether arbitrarily long memory times are achievable as a matter of principle. But we should emphasize that a two-dimensional topologically ordered system of fixed size could be a useful quantum memory if the temperature $T$ is far below the topological gap $\Delta$.

The signatures of topological order, such as quantized Hall conductance and nontrivial topological entanglement entropy, disappear on an $L\times L$ lattice for system size larger than $L_0(T)$, where~\cite{Castelnovo07,Freeman14}
\begin{equation}
\log L_0(T) \approx \Delta /T.
\end{equation}
For $L > L_0(T)$ the system is likely to contain propagating thermally excited anyons, causing rapid decay of stored quantum information. But just as the quantized Hall conductance is a robust physical phenomenon, even though it disappears in the thermodynamic limit at any nonzero temperature, so a topological quantum memory with storage time enhanced by the inverse Boltzmann factor $e^{\Delta/T}$ may be a valuable resource, even though it fails to meet our criteria for self correction.


\subsection{Self correction and fast propagation}\label{subsec:fast-prop}

The Lieb-Robinson bound, which applies to any local Hamiltonian which is a sum of finite-norm terms, asserts that information propagates at finite speed. This bound is violated by the toric-boson model Eq.~(\ref{eq:Pedrocchi-Hamiltonian}); therefore this model cannot arise from an underlying model whose Hamiltonian is a sum of terms, each with bounded norm and bounded range. 

When the toric-boson model on a $L\times L$ lattice is perturbed by a suitable local Hamiltonian, anyons can propagate at speed $O(L)$. Consider dividing the lattice into two domains separated by a horizontal domain wall, as in Fig.~\ref{fig_drop}. The lower half $A$ is in the phase where plaquettes take the value $W_{r}=+1$, and the upper half $B$ is in the phase with $W_r= -1$. At time $t=0$, two anyonic excitations (each with $W_r=+1$) are created deep within the $B$ phase, and at the same time magnetic fields are applied to spins on a vertical line which extends from the lower anyon to the domain wall, and beyond into the $A$ phase. This anyon is subjected to a long-range force attracting it toward the $A$ phase. Were no magnetic field applied, the configuration with an anyon pair would be an exact energy eigenstate and would not evolve. But the applied magnetic field gives the lower anyon a finite effective mass, so it can respond to the attractive force by falling vertically. By the time it reaches the domain wall, this falling anyon has accelerated to speed $O(L)$. Thus anyons can travel at unbounded speed in the limit of infinite system size, showing that the Lieb-Robinson bound does not apply. When the anyon reaches the domain wall, it plunges through, becoming an anyon with $W_r=-1$ in the $A$ phase. Forces attracting it back toward the $B$ phase slow the anyon down until it comes to rest deep with the $A$ phase, and then proceeds to oscillate between positions in the $A$ and $B$ phases. 

\begin{figure}[htb!]
\centering
\includegraphics[width=0.55\linewidth]{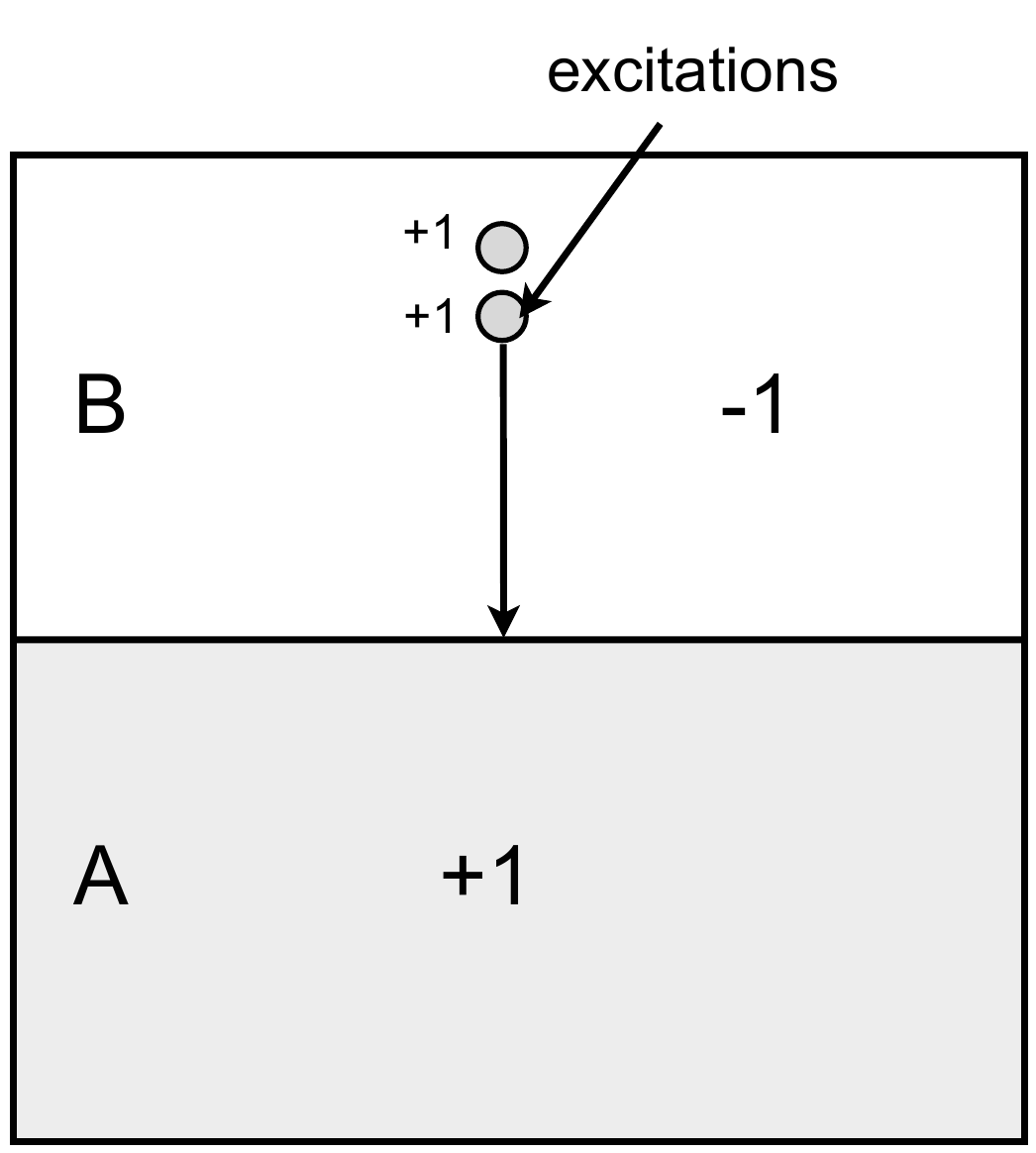}
\caption{A thought experiment to demonstrate the violation of the Lieb-Robinson bound in the toric-boson model. The $A$ phase with $W_{r}=+1$ is separated by a horizontal domain wall from the $B$ phase with $W_r=-1$. An anyon pair is created at time $t=0$ deep within the $B$ phase, and at the same time a magnetic field turns on along a vertical line extending from one anyon to the $A$ phase. This anyon, attracted by the $A$ phase, falls vertically, reaching speed $O(L)$ by the time it reaches the domain wall. 
} 
\label{fig_drop}
\end{figure}

Since the divergent anyon chemical potential of the toric-boson model is directly related to the long-range forces which can accelerate anyons to unbounded speed, one wonders whether this connection holds more generally. Generalized Lieb-Robinson bounds have been derived for systems which have long-range interactions scaling with separation $r$ as $r^{-\alpha}$, showing that in $D$ dimensions the time required for information to propagate a distance $r$ is bounded below by a logarithm of $r$ for $ \alpha > D$~\cite{hastings-math-ph-0507008} and by a sublinear power of $r$ for $\alpha > 2D$~\cite{foss-feig-1410.3466}. No limitations on propagation speed have been derived for $ \alpha \le D$, the regime in which long-range interactions can cause the chemical potential to diverge. Since the infinite anyon chemical potential results from the response of the topological medium arbitrarily far away, it's tempting to assert that a divergent chemical potential can occur only in a system with constant propagation time independent of $r$, though we do not have a rigorous general argument for that conclusion.

Not only bosons, but also fermions can induce long-range interactions when fermion wave functions are delocalized. A familiar example is the RKKY interaction between nuclear magnetic dipole moments induced by exchange of conduction electrons, leading to an effective interaction strength proportional to
\begin{equation}
\sim \frac{1}{r^4} [2kr\cos(2kr) - \sin(2kr)]
\end{equation}
where $r$ is the separation of the dipoles and $k$ is the conduction electron's wave number. Even though such effective long-range forces can arise from fermion exchange, these forces cannot produce an infinite propagation speed or an infinite anyon chemical potential. The microscopic Hamiltonian is a sum of terms with bounded range and bounded norm, so the Lieb-Robinson bound implies a finite propagation speed. Furthermore, the energy cost of creating an isolated anyon is finite, because only a constant number of terms in the Hamiltonian, each with constant norm, are sensitive to the presence of the excitation. Indeed, this remark applies not just to fermions but also to spin models where the Hamiltonian is a sum of terms with bounded range and bounded norm.

\subsection{Self correction as a phase transition}

We expect the stored quantum information in a quantum memory to decay quickly at sufficiently high temperature. Therefore, a system which is self correcting is expected to undergo a phase transition at a critical temperature. In the low-temperature phase the memory time increases without bound as the system size increases, and in the high-temperature phase the memory time is bounded above by a constant independent of system size. If the system has no phase transition at nonzero temperature, then we do not expect it to be self correcting at low temperature. 

For example, though Haah's cubic code~\cite{Haah11} and Michnicki's welded code~\cite{Michnicki12, Michnicki14} both have divergent energy barriers, in both cases the partition function can be computed exactly, and one finds that singularities occur only at zero temperature. The absence of a phase transition at nonzero temperature is compatible with the observation that the memory time is constant at nonzero temperature, despite the increasing energy barrier. In contrast, the four-dimensional toric code, which is self correcting, has a phase transition at nonzero temperature associated with condensation of string excitations \cite{creutz1979PhysRevD}. 

In fact the connection between self correction and phase transitions can be rather subtle, as Hastings \emph{et al.} observed~\cite{Hastings14}. They noticed that in the toric code in six or more dimensions, the critical temperature $T_c$ associated with thermodynamic singularities is higher than the percolation temperature $T_p$, above which thermally excited extended defects can have infinite size with nonnegligible probability; furthermore, the separation between the two temperatures grows parametrically with increasing dimension $D$. A related phenomenon occurs if we consider the toric code in $D=4$ dimensions, but where the variables residing on lattice plaquettes take values in $Z_N$ rather than $Z_2$. As has been known for some time, for $N\ge 5$ the four-dimensional $Z_N$ gauge theory has two phase transitions as the temperature varies \cite{creutz1979PhysRevD}. In the low-temperature phase, defects are typically small and dilute, and the memory time increases exponentially with system size. In the high-temperature phase, stored information decays in constant time. 

To gain insight into the memory time in the intermediate-temperature region, it is helpful to consider classical spin systems storing classical information which also exhibit defect percolation below the critical temperature. For example, for the $N$-state Potts model~\cite{Wu82} with $N\ge 5$, the defect percolation temperature $T_p$, above which thermally fluctuating domain walls have unbounded size, is less than the critical temperature $T_c$, above which the system does not spontaneously magnetize. In the intermediate-temperature phase, although defects percolate, the free-energy barrier for logical errors is $O(\log L)$, and correspondingly the memory time scales polynomially with system size~\cite{Beni14}. This physical picture suggests that the memory time may also scale polynomially for high-dimensional toric codes, when the temperature is in between the percolation temperature and the critical temperature. Thus, while $T < T_c$ may be necessary for an exponentially scaling memory time, it does not seem to be sufficient in general.

Hastings \emph{et al.} also suggested that high-dimensional toric codes may be self correcting even in the superheated regime above the critical temperature, due to hysteretic behavior ~\cite{Hastings14}. Mean-field theory predicts that the Potts model is a good classical memory even for $T > T_c$~\cite{Cuff12}, but this mean-field prediction is not supported by numerical simulations. The memory time of the (slightly) superheated two-dimensional Ising model increases polynomially with system size $L$ up to an optimal size $L^*$ comparable to the correlation length, but the $N$-state Potts model has a first-order thermal phase transition for $N\ge 5$, and its stored classical information decays rapidly when the temperature is just slightly above the critical temperature. 

Another connection between self correction and phase transitions can be obtained from the analysis of master equations describing thermally fluctuating systems. For any classical spin system subject to Glauber dynamics, in which \emph{all} correlations functions decay exponentially with distance, if the Gibbs state is the unique fixed point then the convergence to the Gibbs state is rapid~\cite{Martinelli}. Here ``rapid'' means scaling polynomially with system size; for some classical spin systems stronger results can be derived~\cite{kastoryano2013quantum}, and in any case the classical memory time may in some cases be much shorter than the convergence time to the Gibbs state. Since exponentially decaying correlations are a typical feature of high-temperature phases, we may expect rapid mixing to the Gibbs state above the critical temperature, and therefore rapid decay of stored classical information.




\subsection{Perturbative origin of long-range forces}\label{sec:perturbation}

In Sec.~\ref{sec:review} we discussed how long-range interactions between plaquette operators might enhance the quantum memory time of a two-dimensional topologically ordered system. But until now we have not considered in detail how such long-range forces might arise from a microscopic Hamiltonian which is a sum of terms with bounded range and bounded norm. 

Before we discuss the microscopic origin of long-range interactions, let's first recall how topological order can arise in a two-dimensional spin system. The toric code provides a beautiful and instructive example of a frustration-free commuting Hamiltonian with a topologically ordered ground state. But each term in the Hamiltonian acts nontrivially on four qubits, while the naturally occurring interactions in a spin Hamiltonian are typically two-body terms. In fact, frustration-free commuting models realizing any doubled (non-chiral) anyon model can be constructed \cite{levin2005string}, but only by including even higher-weight many-body terms in the Hamiltonian. What spin models with physically plausible two-local interactions will exhibit topological order?

It is known that the ground state of any two-body frustration-free commuting Hamiltonian has only short-range entanglement, and therefore cannot be topologically ordered~\cite{bravyi2003commutative,Bravyi06}. This conclusion also applies to frustration-free commuting Hamiltonians with three-local interactions among qubits on any graph, or for three-local interactions among qutrits on a nearly Euclidean lattice~\cite{aharonov2011complexity}. In this sense the toric code, with four-local interactions among qubits on a square lattice, is the optimal model of two-dimensional topological order based on an exactly solvable commuting Hamiltonian. 

Therefore, to obtain a topologically ordered ground state from a two-local Hamiltonian, we must be willing to consider noncommuting Hamiltonians, which may not be exactly solvable. One approach to constructing such models is the method of perturbative gadgets~\cite{kempe2006complexity,Kitaev06b}. The idea is to approximate a target Hamiltonian $H_0$, which is commuting and has a topologically ordered ground state, with a two-local Hamiltonian $H$. We may express $H$ as
\begin{equation}
H = H_{0} + V;
\end{equation}
because the topological order of $H_0$ is perturbatively stable, the quantum memory properties of the two-local Hamiltonian $H$ will mimic those of the target Hamiltonian $H_0$ if the perturbation $V$ is a sum of sufficiently small local terms.

But can we also obtain long-range plaquette-plaquette interactions of the form $-\sum_{r,r'} J_{r,r'}W_{r}W_{r'}$ as a perturbative approximation to a two-body Hamiltonian? On a square lattice, each term in this Hamiltonian involves eight lattice sites, and the number of terms scales like the square of the system's area, where each term involves the exchange of at least one gadget particle, so the ``density'' of gadget particles per unit area is divergent. We have not found a way for the gadget particles to couple to the plaquette variables in the desired way, or for the nonlocal Hamiltonian to be perturbatively close to a two-local one. 

In fact, as already noted in Sec.~\ref{subsec:fast-prop}, we do not expect the infinite anyon chemical potential of the toric-boson model to be realizable by any Hamiltonian which is a sum of terms with bounded range and bounded norm, simply because the energy cost of a localized excitation should be finite if the excitation occupies a finite lattice volume. Or at any rate, the toric-boson model could provide a good description only in a scaling limit where the lattice spacing becomes small compared to the anyon size. 

A further potential problem is that long-range interactions, whatever their origin, might interfere with the perturbative stability of the target Hamiltonian, and hence invalidate the perturbative gadget method for achieving topological order. Arguments for the perturbative stability of topological order formulated in~\cite{Bravyi10b,MZ13} do not apply if the perturbation $V$ includes terms which decay algebraically (rather than exponentially) with distance. Indeed, systems with long-range interactions can often be well described using mean-field approximations in which entanglement is assumed to be short range. If a topologically ordered system with long-range plaquette-plaquette interactions can arise from a two-local Hamiltonian at all, approximation schemes going beyond conventional perturbation theory may be needed to analyze the stability of the system's stored quantum information~\cite{Ocko11b}.  

\acknowledgments

We thank Jeongwan Haah, Michael J. Kastoryano, Daniel Loss, Kamil Michnicki, Fernando Pastawski, Fabio Pedrocchi, and Kristan Temme for helpful discussions. BY is supported by the David and Ellen Lee Postdoctoral fellowship. OLC is partially supported by Fonds de Recherche Qu\'ebec-Nature et Technologies. DP is partially supported by Canada's NSERC and the Canadian Institute for Advanced Research. This work was supported in part by NSA/ARO grant W911NF-09-1-0442, and AFOSR/DARPA grant FA8750-12-2-0308. We also acknowledge funding provided by the Institute for Quantum Information and Matter, an NSF Physics Frontiers Center with support of the Gordon and Betty Moore Foundation (NSF Grants No. PHY-0803371 and PHY-1125565).  Part of this work was done while DP was visiting IQIM.

\bibliography{myref_nyaan_2015}

\end{document}